# TIME DELAY ESTIMATION, USING CORRELATION APPROACHES APPLIED TO SEISMIC TIME PICKING

[1] Ashraf H. Yahia, [2] El-Sayed El-Dahshan and [3] Albert K. Guirges

[1&2] Physics Department, Faculty of Science, Ain Shams University, Cairo, Egypt, [3] Basic Science Department, Faculty of Engineering, British University in Egypt, Helwan, Egypt

*Abstract*

In wave propagation theories, many problems of multi-sensor systems utilize time delay in their solution in signal processing. This technique finds great utility in seismic exploration and static correction (low velocity weathering), which compensates the difference in elevations of the surveyed land. Traditionally, cross correlation approaches; such as phase delay, coherence ratio and higher order for instance bispectral techniques are the preferred methods of time delay estimation. In this work, we study the reliability of these approaches for estimating the time delay and proposed an interactive algorithm, which used the estimated time delay for automatically obtaining the first break times of seismic data signals; this is considered an essential step in static correlation stage. Here we show that, the phase delay and coherence ratio are almost equivalent with the higher order (bispectral) correlation technique and take a computational time less than the higher order (bispectral) correlation, so is recommend to use.

## 1. INTRODUCTION

Time delay estimation (TDE) among the received signals, that acquired by different groups of sensors; is essential to solve many of signal processing problems in systems, that transmit sound waves to propagate through unidentified medium and received again with delay, which gives information about this medium. Also, it is used as essential measurement in many applications, such as aircraft detection, weather forecasting, medical imaging and earth mapping.

Several techniques are used to estimate the time delay; they are summarized in [1]. Based on that, the cross correlation (CC) technique is used as a function of the similarity between two signals by correlating both to each other with different shifts and looking for the shift of largest amplitude (maximum similarity) among the CC coefficients, which corresponds to the minimum time delay. On the other hand, due to the nonlinearity and low S/N ratio of seismic data set, the CC is not obligatory to estimate a precise time delay. A generalized cross correlation (GCC) is suggested to convolve a pre-filter, for instance Roth, Scot, Phat and Hannan-Thompson with the CC pattern to make it more better [2-5]. Hilbert peak envelope, using Hilbert transform is utilized to derive an analytical envelope of the CC pattern to overcome the rapid oscillations in the CC pattern and makes the difference between the heights of peaks large enough to transfer the largest peak into CC pattern well [6]. Interpolation technique (parabolic Interpolation) may be added to increase the accuracy of the estimation process by obtaining subsamples of the time delay [7]. These techniques are biased by the source noises, which could result in application problems; for instance seismic data processing where the signals are non-Gaussian processes contrary to noises. Higher order (bispectral) correlation is preferred to use due to its ability to suppress Gaussian processes (noises), since theoretically all polyspectra of Gaussian processes of order greater than two vanish [8].

In this work, we present a set of alternative approaches of time delay estimators, based on the correlation technique to enhance the capability of the estimation process of the time delay. Also, proposing an interactive algorithm for automatically computing the first break of seismic data signals which is a vital step in seismic exploration, based on the estimated time delay.

The organization of this article is going as follow: section II shows a set of approaches of time delay estimators, based on the correlation technique; an algorithm of automatic time picking is introduced in section III; section IV reports the time delay and automatic picking results; and section V covers the conclusion of this work.

## 2. TIME DELAY ESTIMATORS

A time delay estimators are mathematical techniques used to estimate the time delay and through the CC technique; and their approaches are used between two discrete signals in the time domain; these described as reference signal $x(n)$ and $y(n)$ as shifted/delayed signal. It correlates both of them by shifting $y(n)$ signal relative to $x(n)$ signal and finds out the shift corresponding to the maximum similarity factor between two signals. The idea of time delay is clarified from the autocorrelation technique (AC), which known as a particular approach of CC technique, where the signal compared with itself and always its maximum value is located at zero shift. So, the minimum time delay equals to the shift that satisfies the maximum value (peak) among the CC coefficients.

These CC & AC techniques are given in time domain as:

$$CC_{xy}(\tau) = E[x(n)y(n+\tau)] \quad (1)$$

$$AC_{xy}(\tau) = E[x(n)x(n+\tau)] \quad (2)$$

where; E{*} denotes as the expectation operation and $\tau$ is lag.

CC/AC techniques can also be computed in the frequency domain using Parseval's theorem, who stated that "the sum of the square of a signal in time domain equals to the sum of the square of its transform frequency domain", so the complicated convolution process of CC/AC in the time domain turned into just multiplication process in the frequency domain, where the CC & AC techniques expressed in the frequency domain as:

$$CC_{xy}(\lambda) = X(\lambda) \times Y(\lambda)^* \quad (3)$$

$$AC_{xy}(\lambda) = X(\lambda) \times X(\lambda)^* \quad (4)$$

where; $X(\lambda)$ is the discrete Fourier transform of $x(n)$, $Y(\lambda)$ is the discrete Fourier transform of $y(n)$, ($\times$) is just a multiplication operator and (*) is the complex conjugate operator. Besides that, there is another presentation of the CC / AC technique used in this work, where represent both of them in terms of a magnitude and phase and these are given as:

$$C_{xy}(\lambda) = |C_{xy}(\lambda)| exp^{j\emptyset_{xy}(\lambda)} \quad (5)$$

$$C_{xx}(\lambda) = |C_{xx}(\lambda)| exp^{j\emptyset_{xx}(\lambda)} \quad (6)$$

Because of the presence of source noises, the CC correlates both the signals and noises together, which makes the cross correlation isn't reliable to show a peak at time delay lag position. Then, we have the following proposed alternative approaches of the CC; these can be used as generally different time delay estimation methods in production.

### 2.1. Phase Delay Estimator (PDE)

The main idea of PDE method, as proposed in [Ikelle et al, 1997], is measuring the phase shift between the CC and AC techniques. This phase shift could be presented in frequency and it can be expressed as:

$$\partial(\lambda) = \emptyset_{xy}(\lambda) - \emptyset_{xx}(\lambda) \quad (7)$$

where; $\partial(\lambda)$ is the phase shift between them, $\emptyset_{xy}(\lambda)$ is the phase of the CC technique and $\emptyset_{xx}(\lambda)$ is the phase of the AC technique. This phase can be calculated by the ratio between the imaginary and real parts of CC/AC pattern in frequency domain, that may described as:

$$\emptyset_{xy}(\lambda) = tan^{-1}\left(\frac{imaginary(CC_{xy})}{real(CC_{xy})}\right) \quad (8)$$

$$\emptyset_{xx}(\lambda) = tan^{-1}\left(\frac{imaginary(AC_{xx})}{real(AC_{xx})}\right) \quad (9)$$

Then are calculated a new quantity (Q), which is called the coherence ratio,

$$Q(\lambda) = exp^{j\partial(\lambda)} \quad (10)$$

After that, the inverse discrete Fourier transform is applied to get the coherence ratio in time domain as:

$$I(\tau) = \int_{-\pi}^{\pi} Q(\lambda) exp^{-j\lambda\tau} d\lambda \quad (11)$$

As mentioned above, the maximum similarity occurred at lag ($\tau$) equals to the minimum delay between the two signals.

### 2.2. Coherence Ratio Estimator (CRE)

In this approach the coherence ratio is proposed as a normalized version of CC technique by the AC technique in the frequency domain, which is given as:

$$Q(\lambda) = \frac{C_{xy}(\lambda)}{C_{xx}(\lambda)} \quad (12)$$

By substituting in eq. (12) from eqs. (5) & (6) into, the result is:

$$Q(\lambda) = \frac{|C_{xy}(\lambda)|*exp^{j\emptyset_{xy}(\lambda)-j\emptyset_{xx}(\lambda)}}{|C_{xx}(\lambda)|} \quad (13)$$

where, we can approximate that, the two signals have the same power (Homogenous medium):

$$|C_{xx}(\lambda)| \approx |C_{yy}(\lambda)| \quad (14)$$

So, we can write eq. [12] in the present form as;

$$Q(\lambda) = \frac{|C_{xy}(\lambda)|*Q}{\sqrt{|C_{xx}(\lambda)|}\sqrt{|C_{yy}(\lambda)|}} \quad (15)$$

Then continue to obtain the time delay as PDE.

The coherence ratio takes the two values 0 or 1, in which 0 means completely out of phase and 1 means completely in phase. On the other hand, the coherence ratio has a great effect on the source noises in seismic data due to the normalization process.

### 2.3. Higher order correlation estimator (HOCE)

This estimator is introduced by [9] as correlation technique, but in the 3rd order domain, that called the Bispectral correlation (BC), which is used due to its ability to suppress the Gaussian processes (noises) [8] and can be expressed mathematically as,

$$B_{xyx}(\lambda_1, \lambda_2) = E[X(\lambda_1)Y(\lambda_2)X^*(\lambda_1 + \lambda_2)] \quad (16)$$

Similar to the CC technique, a normalized version of the BC technique is introduced called the bicoherence ratio that mathematically defined as:

$$BCR_{xyx}(\lambda_1, \lambda_2) = \frac{|B_{xyx}(\lambda_1,\lambda_2)| * \exp^{\delta'(\lambda_1,\lambda_2)}}{\sqrt{|B_{xxx}(\lambda_1,\lambda_2)|}\sqrt{|B_{yyy}(\lambda_1,\lambda_2)|}} \quad (17)$$

where:

$$\delta'(\lambda_1, \lambda_2) = \phi_{xyx}(\lambda_1, \lambda_2) - \phi_{xxx}(\lambda_1, \lambda_2) \quad (18)$$

The BC is able to correlate and detect the phase shift between two different frequency components of the same signal, but the CC cannot. So, when the frequency component at $(\lambda_1 + \lambda_2)$ is coupled of components at $(\lambda_1)$ and $(\lambda_2)$, the bicoherence ratio takes the two values 0 or 1, in which 0 means completely out of phase and 1 means completely in phase.

### 3. AUTOMATIC TIME PICKING

Generally, some problems in seismic exploration are solved by knowing the travel times, of arrivals; these problems appear in all seismic processing stages, for instance statics correction and their jobs because the first arrivals (picks) distorted with noises. It is useful to extract the times of the signals of seismic data.

We here present an interactive algorithm for the automated picking of the first break times, from a seismic data set, based on the above-mentioned time delay estimators.

The initial parameters to our picking are (1) a reference signal, (2) a reference time, (3) a hypothetical velocity model and (4) a window length. It works on both sides of the data separately to minimize the accumulative error through the picking process as follow:

We correlate between the reference signal with its neighbouring signal by using one of the CC approaches; for instance phase delay, coherence ratio and higher order (bispectral) techniques, then calculating the time delay between them. Next considered the previous neighbouring signal, as the new reference trace and updating the reference time by adding it with the estimated delay. Thus, the whole signals in the data can be picked in similar manner, so we propose the following plan, which has these steps:

- Divide the seismic data set into segments, depending on the nature of data, which minimizes the accumulative error percent.

- Select one reference trace and obtain its reference time (to) for every segment.
- Apply a rectangular time window with a length (gate) and a hypothetical velocity model (v) for every segment to centralize the significant area of signals to pick their times, so we are interesting of the first breaks (refraction energy).
- For every segment, correlate the reference trace with its neighbouring trace, then find out the shift, that correspond to the largest amplitude of the correlation pattern.
- Update the reference time (to) by adding it with a time delay found and updates the adjacent current seismic trace, as the next reference trace. Continue until the end trace in the seismic data.
- Automated the time picking for segments can be done by repeating the process for the whole data.

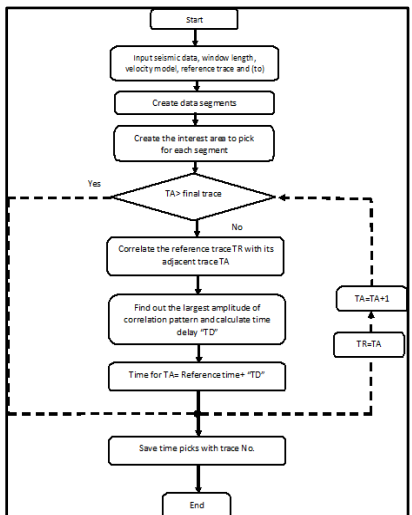

**Figure 1** Flowchart of automatic time picking algorithm

The flowchart of the previous algorithm is shown in Figure 1, which summarizes the previous steps of this algorithm.

## 4. RESULTS AND DISCUSSIONS
### 4.1. Data Set

A real 2D seismic data were extracted from Kansas City for the aim of studying the near-surface seismic reflection data. The acquisition layout was designed as 96 single geophones with 0.5m spacing and the data were recorded for time equals to 500 ms with a sample interval of 0.25 ms [10].

### 4.2. Time Delay Results

To distinguish between our proposed time delay estimators and to show their capabilities of estimation, consider two real discrete seismic signals from Kansas data and correlate between them, using the above-mentioned estimators.

First we windowed these signals with certain length (optimum), these two signals are showed in Figures 2-A and 2-B. The cross correlation, phase delay and coherence ratio approaches calculated between $x(n)$ & $y(n)$ are displayed in Figures 2-C, 2-D and 2-E, respectively. As a result of the nature of signals the CC estimates a biased time delay due to the existence of Gaussian noises with signals However, the phase delay and coherence ratio approaches estimate a good time delay relative to the CC. These are less affected by Gaussian noises, because of normalization processes done.

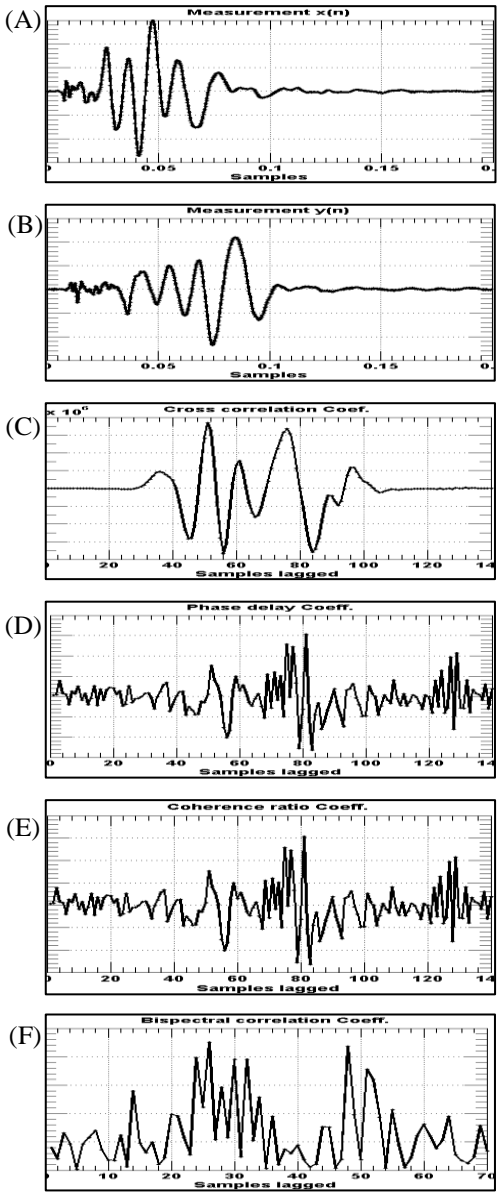

**Figure 1 (A) & (B) two seismic signals, (C) cross correlation coefficients, (D) phase delay coefficients, (E) coherence ratio coefficients and (F) bispectral correlation coefficients.**

The higher order correlation technique (Bispectral correlation) is displayed in Figure 2-F, that shows a less sensitivity to Gaussian noises and it yields an estimate to the time equitable delay. All results of time delay estimation using the above-mentioned estimators are shown in Table 1.

| Type of estimator | Estimated time delay in (s) |
|---|---|
| Manual | 0.0032 |
| Cross correlation | 0.005 |
| Phase delay | 0.0025 |
| Coherence ratio | 0.0025 |
| Bispectral correlation | 0.003 |

**Table 1 A comparison between the estimated time delays using our time delay estimators**

### 4.3. Results Of Automated Time Picking

The picking process starts with just one segment and with certain parameters, which are shortened in the zero-offset trace as a reference trace that has reference time equal to 8.6 ms, a rectangular time window with a length equals 50 ms and a suitable velocity model for the first break event.

The automated time picks for a seismic shot are obtained, based on the above-mentioned estimators; such as cross correlation estimator, phase delay estimator, coherence ratio estimator and higher order correlation estimator (bispectral correlation), and their results of picking are summarized in Figure 3 which shows the automated time picks are overlaid the original seismic data set in red (×) points and displayed the predicted moveouts separately in red (×) on the base of the previous estimators, in which all the proposed pickers show a good picking for the first break times.

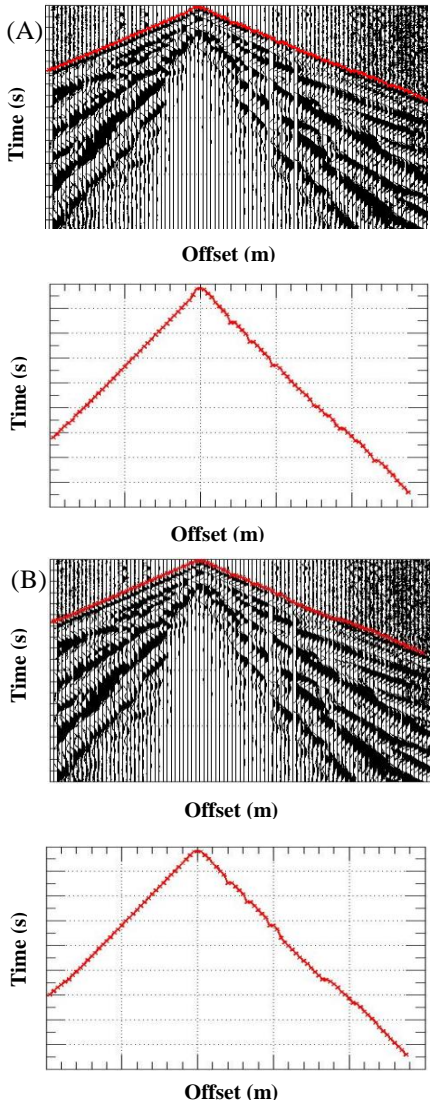

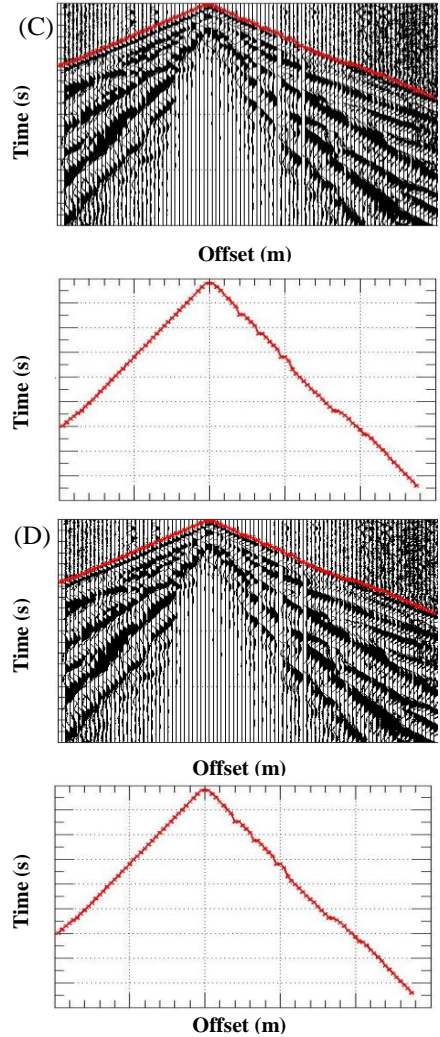

**Figure 2** An example automated time picking is overlaid on the original data by using (A) cross correlation estimator, (B) Phase delay estimator, (C) Coherence ratio estimator and (D) Higher order correlation estimator.

As shown in Figure 3-A, the cross correlation picked biased times that are improved by using both of the PDE and CRE, which have the same picked times presented in Figure 3- B and 3-C, and Figure 3-D. The BCR picked also as the PDE and CRE ratio estimator due to its ability to suppress the Gaussian noises.

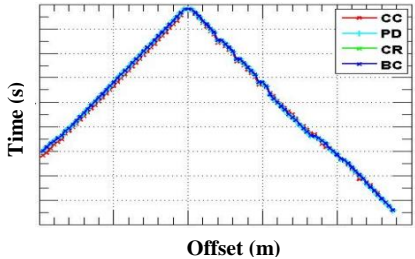

**Figure 3** A comparisons between the automated first break picking using our time delay estimators

Apparently in Figure 4, the automated time picks based on both the PDE and CRE are equivalently to each other and in the same time equivalently to the automated time picks based on BCR. So, we can use one of them to improve the automated time picking instead of the cross correlation technique estimators.

## 5. CONCLUSIONS

In conclusion, we have used our improved time delay estimators, which summarized in CC, PDE, CRE and BCE to estimate the time delay in seismic exploration, and used them to pick the times of first breaks. We find that, the PDE or CRE present, an almost equivalent picks and take computational time less than BCE, so we recommend using one of them in large data set. Also the BCE is not conformed in the production.

## 6. ACKNOWLEDGEMENTS

The authors would like to strongly acknowledge Dr. Larry Velasco, researcher of RD Schlumberger Company on his efforts of this project and Prof. Dr. Ahmed Abdel Atti, professor of geophysics, Faculty of Science, Ain shams University, for his careful reading and reviewing of the manuscript. Also, his useful and fruitful discussions are greatly appreciated.

## 7. REFERENCES

[1]. Carter, 1987. Coherence and time delay estimation. Proc. IEEE, vol. 75, pp. 236-255.

[2]. Knapp, Carter, 1976. The generalized correlation method for estimation of time delay. IEEE Trans. Acoustics., Speech, Signal Processing, vol. ASSP-24, pp. 320-327

[3]. Carter,1981. Time delay estimation for passive sonar signal processing,'' IEEE Trans. Acoust., Speech, Signal Processing, vol.ASSP-29, pp. 463-470.

[4]. IEEE Trans. Acoust., Speech, Signal Processing, 1981 Special Issue on Time Delay Estimation, vol. 29.

[5]. ----, 1981. Special Issue on Time Delay Estimation. IEEE Trans. Acoustics, Speech, and Signal Processing, vol. 29.

[6]. Thrane et al, 1995. Practical use of the Hilbert transform. Brüel and Kjær Sound & Vibration Measurement A/S Application Note

[7]. Maskell , Woods, 1999.The estimation of subsample time delay of arrival in the discrete-time measurement of phase delay. IEEE Trans. Instrum. pp. 1227–1231.

[8]. Nikias , Pan, 1998.Time delay estimation in unknown Gaussian spatially correlated noise. IEEE trans. Acoustic speech signal processing vol. 36, pp.1706-1714

[9]. Ikelle et al, 1997.An example of seismic time picking by third-

order bicoherence. Geophysics journal internet, Vol.62, No. 6, pp. 1947-1951

[10]. Baker, 1999. Processing near – surface seismic-reflection data. primer, SEG, New York